\documentclass{osa-article}

\journal{oe}

\begin{document}

\title{Towards an \textit{in situ}, full-power gauge of the \\ focal-volume intensity of petawatt-class lasers}

\author{C. Z. He,\authormark{1,2} A. Longman,\authormark{3} J. A. P\'erez-Hern\'andez,\authormark{4} \\M. de Marco,\authormark{4}, C. Salgado,\authormark{4} G. Zeraouli,\authormark{4} G. Gatti,\authormark{4}\\ L. Roso,\authormark{4} R. Fedosejevs,\authormark{3} and W. T. Hill, III\authormark{1,2,5,*}
}
\address{\authormark{1}Joint Quantum Institute, University of Maryland, College Park, MD 20742, USA\\
\authormark{2}Institute for Physical Science and Technology, University of Maryland, College Park, MD 20742, USA\\ 
\authormark{3}Electrical and Computer Engineering, University of Alberta, Edmonton, Alberta T6G 2V4, Canada\\
\authormark{4}Centro de L\'aseres Pulsados (CLPU), 37185 Villamayor, Salamanca, Spain\\
\authormark{5}Department of Physics, University of Maryland, College Park, MD 20742, USA\\
}

\email{\authormark{*}wth@umd.edu}



\begin{abstract}
About 50 years ago, Sarachick and Schappert [Phys. Rev. D. \textbf{1}, 2738-2752 (1970)] showed that relativistic Thomson scattering leads to wavelength shifts that are proportional to the laser intensity.  About 28 years later Chen \textit{et al.} [Nature \textbf{396}, 653-655 (1998)] used these shifts to estimate their laser intensity near $10^{18}$ W/cm$^2$.  More recently there have been several theoretical studies aimed at exploiting nonlinear Thomson scattering as a tool for direct measurement of intensities well into the relativistic regime.  We present the first quantitative study of this approach for intensities between $10^{18}$ and  $10^{19}$ W/cm$^2$.  We show that the spectral shifts are in reasonable agreement with estimates of the peak intensity extracted from images of the focal area obtained at reduced power.  Finally, we discuss the viability of the approach, its range of usefulness and how it might be extended to gauge intensities well in excess of $10^{19}$ W/cm$^2$.
\end{abstract}

\section{Introduction}
Since the invention of the laser, physicists have been dreaming of ways to exploit super-intense laser fields to explore new physics in the relativistic regime.  Arguably, the most fundamental idea today centers on the nature of the quantum vacuum, which could hold key information necessary for unraveling the mystery of dark matter \cite{Hill:2017}.  Even with current advances, we are still orders of magnitude away from reaching the threshold intensity for creating electron-positron pairs directly via photon-photon collisions, the non-perturbative production of which is estimated to be $\sim 2 \times 10^{29}$ W/cm$^2$.  Nevertheless, petawatt-class lasers of short duration have placed us on the cusp of being able to examine experimentally, nonlinear aspects of electrodynamics \cite{Tommasini:2009, Dunne:2009, Hill:2017, BrightestLight:2018}, photon-photon interaction and precursor processes to pair production \cite{Halpern:1933, Euler:1935, Euler:1936, Liang:2012, Heinzl:2006, Tommasini:2014}, which comprise tests of quantum electrodynamics (QED) in ways it has never been tested (see, for example \cite{BrightestLight:2018, Blinne:2019} and reference therein).  Precision measurements of QED require accurate knowledge of the intensity.  Measuring the intensity in the focus at relativistic strengths is challenging.  We define the threshold for relativistic strength as $I \sim 10^{18}$ W/cm$^2$, the point at which the motion of free electrons becomes relativistic when exposed to a laser field of this strength.  A measurement technique must not only be able to determine the true intensity correctly, it also should be minimally intrusive and cause no optical damage to instrumentation.  In addition, the approach would ideally be single-shot capable, allowing the intensity to be determined with each laser pulse.  Developing suitable techniques is a long-standing desire of the laser community at large.  A host of other studies from generating secondary sources \cite{Roth:2013} to high-energy density physics \cite{Dyer:2008, Roth:2009, Feldman:2017} to medical applications \cite{Roth:2002, Jones:2005, Lefebvre:2006, Ledingham:2007, Liu:2015, Tommasino:2015} also would benefit from better knowledge of the intensity.  

A number of techniques and proposals for intensity assessment have emerged over the years.  Most, however, rely on approximating the peak intensity with methods requiring the laser to be run at significantly reduced powers.  A one-to-one correspondence between what is extracted in these cases and the actual intensity, has never been established experimentally.  Since the early observation of Ref.\ \cite{Chen:1998}, there have been three proposals for direct-measurement that can be found in the literature, which, in principle, meet the criteria above.  Two are based on relativistic Thomson scattering (RTS) \cite{Sarachick:1970, Castillo-Herrera:1993, Gao:2004, Gao:2006, Har-Shemesh:2012, Tarbox:2015, Harvey:2018} and the third on the appearance intensity for inner-shell tunnel ionization \cite{Ciappina:2019}.  To our knowledge, there have been no reports of experimental verification of any of these approaches for $I > 10^{18}$ W/cm$^2$.  This communication is devoted to addressing that void.  Specifically, we present the first peak intensity assessment using the RTS spectrum significantly above $10^{18}$ W/cm$^2$.  We compare the RTS peak intensity ($I^{\mathrm{RTS}}_\mathrm{pk}$) with estimates extracted from low-power images of the spatial profile of the focal spot ($I^{\mathrm{Im}}_{\mathrm{pk}}$).  Our analysis shows that the spectral and temporal phase must be monitored in the target area to ensure that the Strehl ratio and pulse widths, typically measured in the laser bay, are close to their optimal values after beam transport and pulse compression to maximize the energy in the central Airy spot.

We begin our discussion by briefly outlining the principle of the RTS intensity gauge, which exploits free electrons that are always produced via ionization when $I \gtrsim 10^{14}$ W/cm$^2$.  Electrons undergoing RTS produce a rich spectrum of intensity-dependent, Doppler-shifted laser wavelengths and their harmonics \cite{Castillo-Herrera:1993}.  In a frame moving in the direction of the laser $\vec{k}$ vector (the average drift frame), the electron approximately executes a figure-8 (see Fig.\ 1 in \cite{Sarachick:1970}) while in the lab frame the electron oscillates in the $\vec{E} \vec{k}$-plane while being propelled along $\vec{k}$, punctuated by cusps (see Fig.\ 2 in \cite{Tarbox:2015}) where large acceleration exists.  The benefit of exploiting RTS radiation is threefold:
\begin{itemize}
\item it can be generated from low-density, background gas in the experimental chamber; 
\item it can be captured with off-the-shelf optics near, but outside, the focal volume; and 
\item the Doppler shifts are straightforward to calculate, based on earlier theoretical work \cite{Sarachick:1970, Castillo-Herrera:1993, Gao:2004, Gao:2006}, leading to a simple expression for the intensity as a function of the observed $n^\mathrm{th}$-harmonic wavelength, $\lambda^{(n)}$,
\begin{equation}
	I(\lambda^{(n)}, \theta; n) = \frac{2 \pi m_e c^3}{r_0 \lambda_0^2 (1 - \cos \theta)} \left( \frac{n \lambda^{(n)}}{\lambda_0} - 1 \right), 
	\label{eq:intensity_ds}
\end{equation}
\noindent where $\lambda_0$, $\theta$, $m_e$, $r_0$ and $c$ are the laser wavelength, the polar angle relative to $\vec{k}$, the electron mass, the classical electron radius and the vacuum light speed respectively.   
\end{itemize}

\noindent At relativistic intensities, when the normalized vector potential is of order unity or greater, $\lambda^{(n)}$ differs measurably from $\lambda_0/n$.  The normalized vector potential is defined as
\begin{equation}
	a_0 = e E \lambda_0 / 2 \pi m_e c^2 \approx 0.855 \sqrt{I [\times 10^{18}\ \mathrm{W/cm}^2]} \lambda_0 [\mu \mathrm{m}], 
\end{equation}
\noindent where $e$ is the elementary charge.   In addition, for intensities in excess of $10^{25}$ W/cm$^2$ such that $a_0 (m_e/m_p) \sim 1$, where $m_p$ is the proton mass, the technique might be extendable to an effective proton RTS.  These features make RTS a potentially useful diagnostic tool for a variety of experiments at relativistic intensities.

\section{Experimental method}

\begin{figure}[t]
	\centering
	\includegraphics[width=1\linewidth]{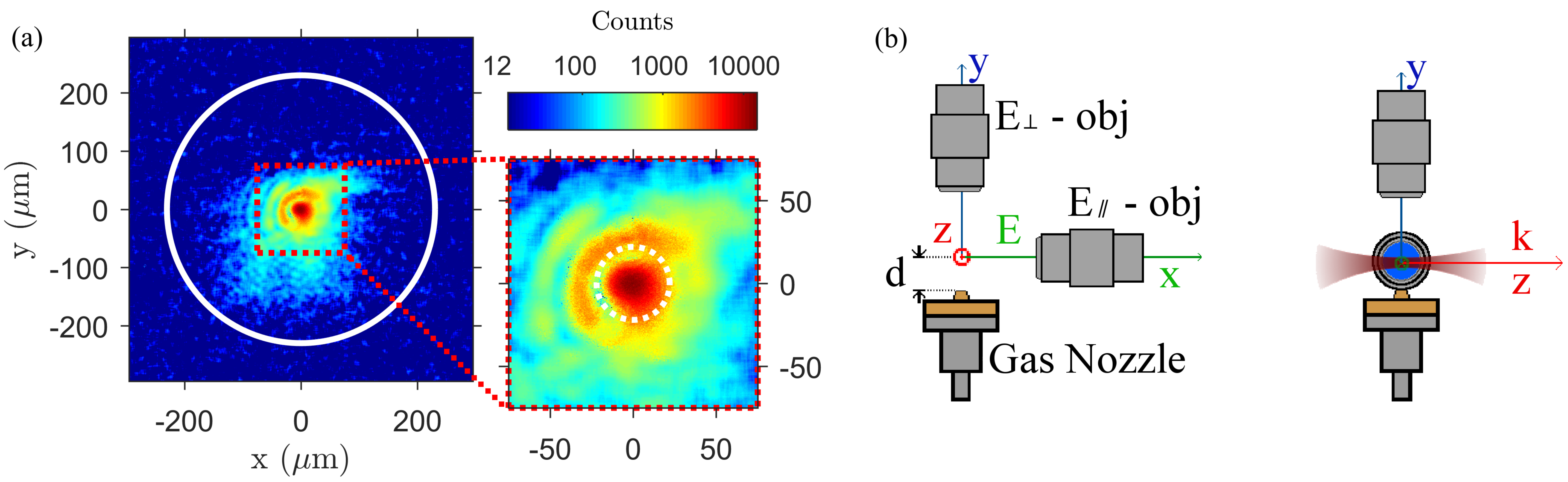}
	\caption{(a) Cross sectional image of the focal volume and an expanded view of the central Airy disk and portions of three secondary rings.  The solid (dashed) white circle indicates the extent of the pulse (central Airy disk).  The signal in each image was smoothed and filtered to remove hot spots and reduce the noise, and clipped at the background count $\sim 12$.  (b) Schematic of the focal region showing the orientation of both microscope objectives ($E_{\perp}$-obj and $E_{\parallel}$-obj), gas nozzle, $\vec{E}$ and $\vec{k}$ vectors.  Note, the $\hat{x}$ and $\hat{y}$ axes correspond to ($\theta, \phi$) = ($\pi/2, 0$) and ($\pi/2, \pi/2$) respectively, with $\theta$ and $\phi$ being the polar and azimuthal  angles respectively. 
	} 
	\label{fig:schematics}
\end{figure}

Our measurements were done at the Centro de L\'aseres Pulsados (CLPU) \cite{CLPU} with a 200 TW Ti:sapphire laser (VEGA 2) running at 10 Hz with $\lambda_0 \sim 790$ nm and compressed to nominally 30 fs.  For our experiments, linearly polarized pulses were used with uncompressed energies between 2 and 6 J.  About 47\% of this energy, the energy on target ($U$), was delivered to the focal volume as compressed pulses after being reflected by an $f/$13, focal length (f.l.) $= 130$ cm, off-axis parabola (OAP).  The diameter (FWHM, i.e., $2 w_0$) of the central Airy peak, measured at low power, was $\sim 22\ \mu$m ($\sim2\ \times$ the diffraction limit) and contained $\sim 33$\% of $U$, i.e., $\sim $16\% of the uncompressed energy.  Figure \ref{fig:schematics} (a) shows an image of the focal spot taken with a microscope objective at low intensity.  

\begin{figure}[t]
	\centering
	\includegraphics[width=1\linewidth]{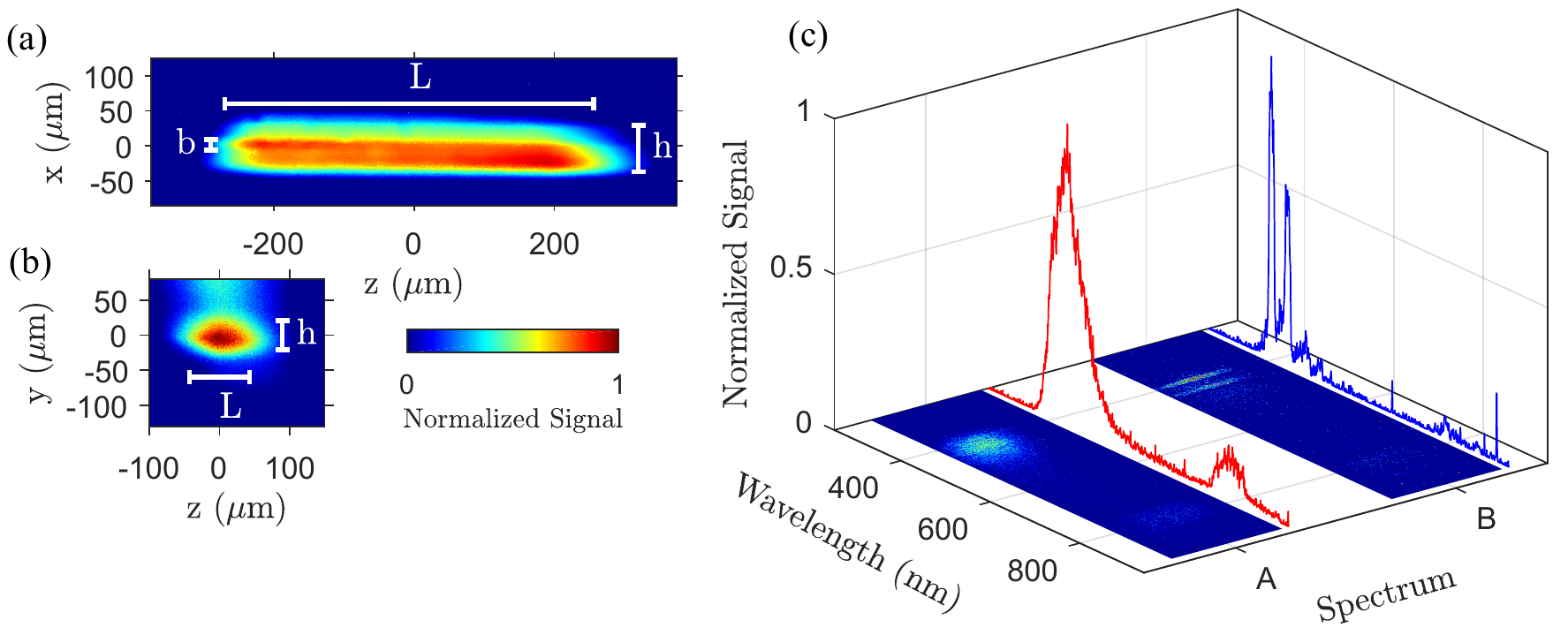}
	\caption{Composite images (500 shots each) obtained with $E_{\perp}$-obj (a) and $E_{\parallel}$-obj (b), for $U = 2.67$ J, which includes both RTS and recombination radiation, where $L$, $h$ and $b$ are the FWHM length $\parallel$ to $\vec{k}$, multi-shot width $\perp$ to $\vec{k}$ and single-shot width $\perp$ to $\vec{k}$ (see text) respectively with values $L= 526/85\ \mu$m, $h= 66/41\ \mu$m and $b=15\ \mu$m$/$N.A. for \mbox{image (a)}/image (b) respectively.  Composite images of the dispersed light and associated integrated profiles (c) for the prompt signal (A), obtained within a 5 ns temporal window that starts $\sim 2.5$ ns before the laser pulse, and the delayed signal (B), obtained within a 50 ns temporal window starting $\sim 2.5$ ns after the laser pulse.  }
	\label{fig:camera-images}
\end{figure}

Electrons were generated from N$_2$ (which yielded up to ten electrons per molecule), delivered to the focal volume by a custom-machined conical nozzle (0.8 mm throat orifice diameter) placed $\sim 1.8$ cm below the focal point, a distance chosen to minimize scattered light from the nozzle while maximizing gas delivered to the interaction region.  The nozzle backing pressure was $10\pm 3$ mbar. The orifice was open for $4$ ms and triggered $\sim2.5$ ms before each pulse.  The pressure in the focal volume was estimated to be $\sim10^{-3}$ mbar, yielding a density of $\sim 2.4\times10^{13}$ cm$^{-3}$, assuming ideal gas behavior.  The focal volume was taken to be a cylindrical column with a diameter equal to the FWHM diameter of the central Airy peak ($\sim 22\ \mu$m).  The effective length of the column was defined by the field of view of the collection optics, $\sim 85\ \mu$m (see text below and Fig.\ \ref{fig:camera-images}), leading to a focal volume of $\sim 3 \times 10^{-8}$ cm$^{3}$.  When the intensity reaches $\sim 10^{16}$ W/cm$^3$, a combination of mechanisms including enhanced ionization \cite{Zuo:1995}, Coulomb explosion (once three or more electrons are removed) and barrier suppression/tunnel (so-called ADK) ionization \cite{Ammosov:1986, Tong:2002} induce N$_2$ to lose all six valence ($p$) electrons leading to N$^{3+}$ ions.  Additional (inner-shell) electrons can be removed via post-dissociation ADK ionization as the intensity rises through $10^{17}$ and $10^{19}$ W/cm$^2$.  

The RTS radiation was captured from two orthogonal directions:  ($\theta, \phi$) = ($\pi/2, 0$) and ($\pi/2, \pi/2$) respectively, see Fig.\ \ref{fig:schematics} (b).  Infinity-corrected microscope objectives (Mitutoyo MLWD-10$\times$ and Motic ELWD-10$\times$) were used, each having an effective $f.l.=20$ mm and numerical aperture $0.28$, corresponding to a half angle for light collection of $\sim 16^\circ$.   The objectives were adjusted with mirror mounts placed on stepper-motor controlled xyz stages.  Preliminary alignment of the objectives was enabled by scatting HeNe and diode beams, collinear with the VEGA-2 beam, off of a needle placed at the intended OAP focus.  Next, a plasma column was generated by running VEGA~2 at low energy, with the focal volume filled with a few tens of millibars of N$_2$.  The emission column was then focused onto the detectors by adjusting both the OAP and the objectives position.  Final adjustments were done by optimizing the RTS signal at full power.  Light captured by the objectives was directed along light-shielded paths (inside and outside the chamber) through a viewport to be analyzed.  A second component of the scattered-light remediation was a beam dump housed in a separate chamber $\sim 1$ m after the focal volume, consisting of flat-black aluminum foil oriented at 45$^\circ$ with respect to $\vec{k}$ and enclosed by two coaxial cylindrical walls (flat-black foil) with an entrance hole just large enough for the laser to pass unobstructed.  

We detected $E_{\parallel}$-obj light either directly with a cooled, low-noise CMOS camera (Zyla 4.2MPPLUS) or by a gateable (5 ns resolution) intensified CCD camera (Andor iStar DH 720-25U-03 ICCD) after being dispersed by a $0.25$-m spectrometer (Oriel MS260i).  The desired detector was selected with a flip mirror.  We detected $E_{\perp}$-obj light either with a cooled, back-side thinned CCD camera (Andor DV440-BU2-2492) or by the same spectrometer-CCD combination just mentioned; selection was again mediated by a flip mirror.  Lenses with $f.l. = 300$ mm were used to create the images on the cameras/spectrometer, giving an approximate effective magnification of 15.  A Schott BG39 filter \cite{Schott}, with a transmittance $> 0.1$ for $320 < \lambda < 660$ nm and $< 10^{-4}$ for $\lambda < 320$ nm and $> 780$ nm (see Fig.\ \ref{fig:sidespectrumset-and-intensityvsenergy} (a) bottom trace), was placed in the path before the detectors to minimize the laser light contribution to the images.   

Data was collected in 10 s bursts (100 shots/burst); the bursts were repeated multiple times (typically five) to create a shot batch.  Each shot batch, obtained with the nozzle open, was followed by a second series with the nozzle closed -- a background batch.  Composite images were generated by combining the two batches:  (shot batch) - (background batch).  The experimental data set consisted of composite images obtained under different conditions.  Specifically, we varied the pulse energy and chirp, the optical filters (i.e., no filter, with BG39 filter or with RG850 filter), the temporal collection window for the Andor iStar CCD camera and the target pressure.  Examples of composite images of the focal region, 500 shots without spectral dispersion, in the 2$^{\mathrm{nd}}$ harmonic ($n=2$) spectral region, are shown in Fig.\ \ref{fig:camera-images} (a) and (b).  Examples of dispersed composite images for the same spectral region are shown in Fig.\ \ref{fig:camera-images} (c).      

To facilitate a comparison between the direct RTS determination of the intensity and an indirect approach, we will analyze the focal spot image (Fig. \ref{fig:camera-images} (a)).  We captured this low-intensity image by reflecting a small portion of the beam, with a wedge beamsplitter after the OAP, into a third infinity-corrected microscope objective placed near the focus of the reflected beam.  We detected the magnified image of the focal spot with a Blackfly PGE 23S6C CMOS camera (having square pixels with 5.86 $\mu$m sides) after reducing the intensity further with neutral density filters.  We imaged the spot size at a series of locations around the focus by adjusting the position of the microscope objective with a translation stage.  We calibrated the Blackfly magnification by measuring the experimental focal spot directly (without the beamsplitter) with a Chamelon CCD camera (having square pixels with 3.75 $\mu$m sides) leading to $0.51 \pm 0.04\ \mu$m/pixel for the Blackfly camera.  

\section{Results and analysis}

\begin{figure}[t]
	\centering
	\includegraphics[width=1\linewidth]{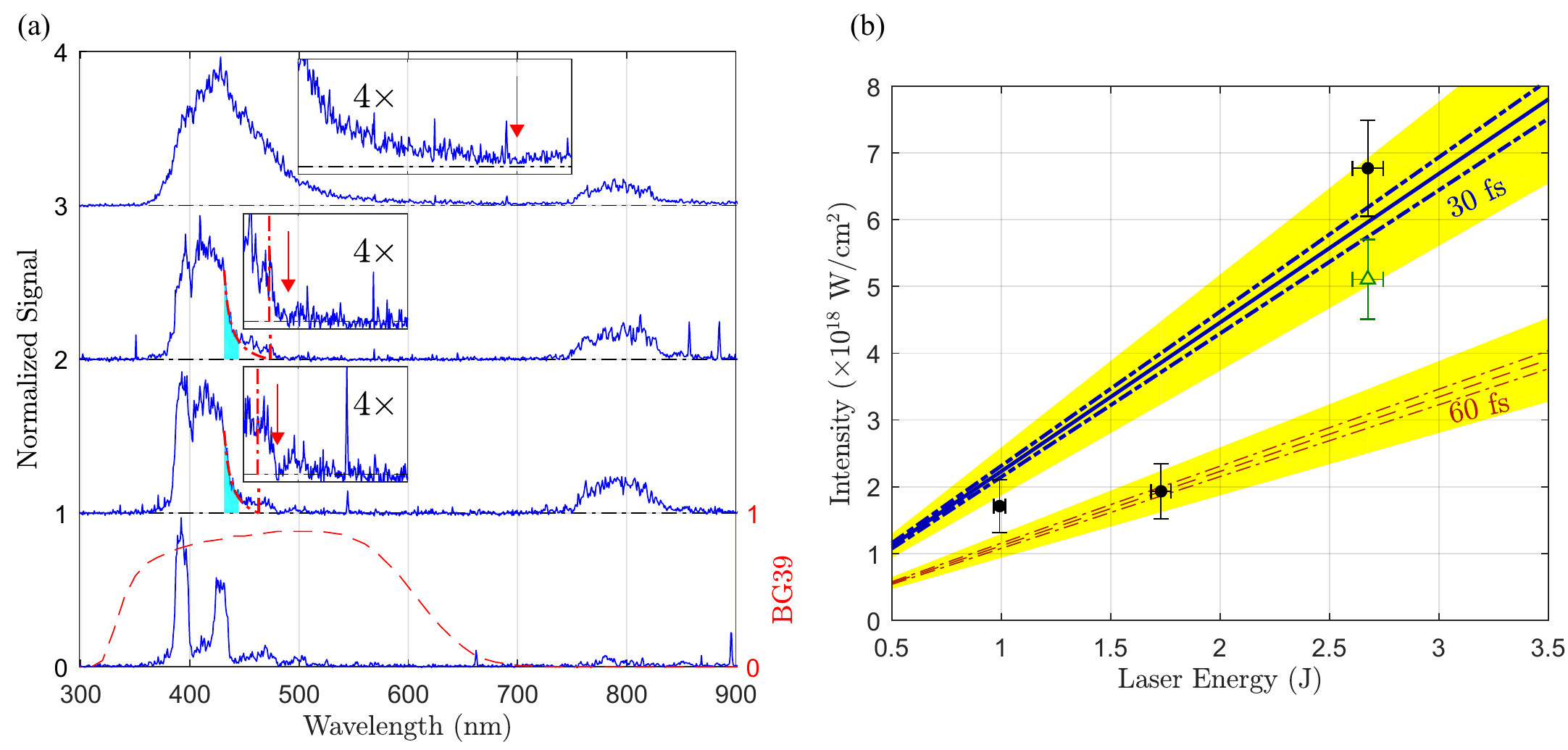}
	\caption{(a) Respectively from the top, RTS-spectral profiles (solid curves) captured by $E_{\parallel}$-obj for three values of $U$ ($2.67 \pm 0.08,\ 1.73 \pm 0.05,\ \mathrm{and}\ 0.99 \pm 0.03$ J) and the plasma recombination lines (solid curve) along with the BG39 filter transmission (dashed red curve).  The red arrows in the insets indicate $\lambda^{(2)}_{\mathrm{r}}$ as determined when the spectrum first reaches $\mathrm{Sig_{bk}}$ while the dot-dash line indicates $\lambda^{(2)}_{\mathrm{r}}$ from the numerical fit to the shaded regions (see text).  (b) Comparison between $I_{\mathrm{pk}}^{\mathrm{RTS}}$ for the $U$ values (black solid circles) and $I_{\mathrm{pk}}^{\mathrm{Im}}$ for $\Delta t = 30$ fs (blue solid line) and 60 fs (red dashed line).  The shaded areas are the uncertainties in $I_{\mathrm{pk}}^{\mathrm{Im}}$ ($\sim$16.1\%); the dot-dashed lines are the uncertainties in $I_{\mathrm{pk}}^{\mathrm{Im}}$ if $A_{\mathrm{pix}}$ were known exactly ($\sim$3.6\%, see text).  The open green triangle $I_{\mathrm{pk}}^{\mathrm{RTS}}$ point was taken at $U=2.67$ J but after the low energy points (see text).
	} \label{fig:sidespectrumset-and-intensityvsenergy}
\end{figure}

Before we discuss the analysis and how we can extract the peak intensity from composite images and Eq.\ \ref{eq:intensity_ds}, there are a few salient observations we can make.  First, Fig.\ \ref{fig:camera-images} (c) makes clear the need to gate the signal in order to isolate RTS and plasma recombination signals.  The RTS radiation is prompt, only occurring during the laser pulse, while the plasma radiation begins a few tens of nanoseconds later and can be substantially stronger.  Gating not only helps to minimize the recombination signal, it serves to block any laser light scattered from the chamber walls.  This emphasizes the need for employing a large chamber so that the walls are far from the collection optics.  It is clear when comparing the two signals in Fig.~\ref{fig:camera-images} (a) and (b), that only the central portion of the radiation captured by $E_{\parallel}$-obj makes it to the detector; the outer portion was  clipped along the transfer path.  We note that the weak signal above the primary signal in Fig.\ \ref{fig:camera-images} (b) is due to residual, off-axis scattered light.  To ensure $E_{\parallel}$-obj was viewing the focal spot, this objective was adjusted to maximize the on-axis signal.  The signal from $E_{\perp}$-obj makes clear $h$ is composed of multiple thinner lines of widths $\sim 15\ \mu$m, closer to the anticipated focal width.  The multiple lines are most likely caused by pointing instability on the order of 1 to 2 beam diameters; this instability appears to occur in both horizontal and vertical directions.

Turning our attention to estimating the RTS intensity, we focus more closely on the prompt spectrum displayed in Fig.\ \ref{fig:camera-images} (c).  This spectrum contains information about the peak intensity and spatial intensity profile of the pulse, both of which are extractable, in principle, from Eq.\ \ref{eq:intensity_ds}.  Determining the intensity profile, however, requires detailed knowledge of when in time and where in space each electron contributing to the measured RTS radiation is born (i.e., knowledge of the appearance intensities such as discussed in Ref.\ \cite{Ciappina:2019}), its trajectory induced by the field, and a reliable numerical RTS model providing quantitative information about its contribution to the captured light in various wavelength bands.  With such, a functional fit of the data to Eq.\ \ref{eq:intensity_ds} would be possible.  Currently, a suitable model does not exist; however, this team is actively working on this problem.  As a first step in developing a diagnostic tool, we will focus on the RTS peak intensity, $I_{\mathrm{pk}}^{\mathrm{RTS}}$, which corresponds to the red spectral edge, $\lambda^{(n)}_{\mathrm{r}}$, in the prompt spectrum.  We will confine our attention to $n = 2$.  Because the $n=2$ signal captured by $E_{\parallel}$-obj is larger than that captured by $E_{\perp}$-obj, as expected (see Fig.\ 5 in \cite{Paredes:2012}), we will focus on composite images from $E_{\parallel}$-obj in the remainder of this manuscript.     

To that end, we display the prompt spectra for three $U$ values ($0.99 \pm 0.03,\ 1.73 \pm 0.05\ \mathrm{and}\ 2.67\pm 0.08$ J) in Fig.\ \ref{fig:sidespectrumset-and-intensityvsenergy} (a).  We take $\lambda^{(2)}_\mathrm{r}$ to be the point where the spectrum falls to the background, $\mathrm{Sig_{bk}}$.  For $U = 2.67$ J (Fig.\ \ref{fig:sidespectrumset-and-intensityvsenergy} (a) top), $\lambda^{(2)}_{\mathrm{r}} \sim 700$ nm.  The values for $\lambda^{(2)}_\mathrm{r}$ are less clear for the other two energies, because of interference from the plasma recombination radiation.  For these energies, we estimated $\lambda^{(2)}_\mathrm{r}$ two ways:  (i) from the location where the curve falls to $\mathrm{Sig_{bk}}$ and (ii) from a numerical fit of the tail of the distribution, identified by the shaded section, to a rational  polynomial, $f(\lambda) = a + b/(\lambda + c)$ with $a$, $b$ and $c$ being fitting parameters.  In the latter we solve $f(\lambda) = \mathrm{Sig_{bk}}$ to determine $\lambda^{(2)}_\mathrm{r}$.  This is also complicated by the presence of the plasma recombination lines distorting the shape, which is likely to cause the fit curve to decay to $\mathrm{Sig_{bk}}$ prematurely.  Because of the plasma contribution, the first method may select a $\lambda^{(2)}_\mathrm{r}$ that is too large at low intensities.  Our estimates of $\lambda_\mathrm{r}^{(2)}$ for these intensities are averages of the two values.  Even though $\lambda_r^{(2)}$ is clear in the top trace of Fig. \ref{fig:sidespectrumset-and-intensityvsenergy} (a), $f(\lambda)\approx \mathrm{Sig_{bk}}$ at a wavelength where the BG39 transmission is $\sim 1$\% and falling sharply.  Thus, $\lambda_r^{(2)}\sim700$ nm may be a lower limit of the red shift.  (We note that a numerical fit to the tail of the $U = 2.67$ J plot also gives $\lambda_r^{(2)}\sim700$ nm.) Using $\lambda_0 = 790$ nm, obtained from a Wizzler \cite{Wizzler} pulse characterization, we estimate $I_{\mathrm{pk}}^{\mathrm{RTS}}(\lambda^{(2)}_\mathrm{r}) = (1.71 \pm 0.40$, $1.94 \pm 0.41$, $6.77 \pm 0.72) \times 10^{18}$ W/cm$^2$ from lowest to highest $U$.  The uncertainty in $I_{\mathrm{pk}}^{\mathrm{RTS}}$ is based on half of the wavelength spread ($\sim39/2$ nm $= 19.5$ nm) of the laser, while the uncertainty in $U$ is based on the RMS energy fluctuations of $\sim$2.65\%.

We now compare the three $I_{\mathrm{pk}}^{\mathrm{RTS}}$ values with $I_{\mathrm{pk}}^{\mathrm{Im}}$ values obtained from Fig.\ \ref{fig:schematics} (a).  There are three basic requirements for a good intensity estimate from such an image.  First, to prevent image saturation so that the intensity profile can be measured over many decades, the camera must have a large dynamic range.  Second, because the optical path contains lenses, windows and filters causing unknown magnification, the physical size of the camera pixels must be calibrated as mentioned earlier.  Finally, assuming the spatial and temporal parts of the intensity profile are separable, an integrable temporal profile function is needed.   When these conditions are met it is straightforward to show that the intensity associated with the energy $U$ on target is given by 
\begin{equation}
	I_{\mathrm{pk}}^{\mathrm{Im}}=\frac{UC_{\mathrm{pk}}}{T_{\mathrm{eff}} C_{\mathrm{sum}} A_{\mathrm{pix}}},
	\label{eq:image-intensity}
\end{equation}
\noindent where $C_{\mathrm{pk}}$, $\ T_{\mathrm{eff}}$, $C_{\mathrm{sum}}$ and $A_{\mathrm{pix}}$ are, respectively, the count associated with the peak intensity, an effective pulse width, the total count summed of the pixels comprising the entire pulse and the area of a camera pixel in physical units.  We derive this equation in the appendix.  The value attached to $C_{\mathrm{pk}}$ is the average value of nine pixels in a $3\times3$ block of pixels near the the peak of the pulse on the image in Fig. \ref{fig:schematics} (a).  To estimate $T_{\mathrm{eff}}$, we first integrated the Wizzler pulse profile, which was approximately Gaussian; however, the Wizzler measurement was made in the laser bay before beam transport to the target area.  Because the pulse width estimated by the Wizzler ($\Delta t_{\mathrm{W}}$) was smaller than that estimated by an autocorrelator ($\Delta t_{\mathrm{AC}}$) in the target area by $\sim$20\% (e.g., $\Delta t_{\mathrm{W}}=25.2\pm0.4$ fs and $\Delta t_{\mathrm{AC}}=31.0 \pm 0.4$ fs), we rescaled $T_{\mathrm{eff}}$ by the ratio $\Delta t/\Delta t_{\mathrm{W}}$, where $\Delta t$ is a nominal pulse width of $30$ fs or $60$ fs in this case.  The value of $C_{\mathrm{sum}}$ is obtained by summing the pixels on the image that comprise the pulse (i.e., inside the white circle in Fig. \ref{fig:camera-images} (a)).  From the magnification calibration, $A_{\mathrm{pix}}=0.26\pm0.04\ \mu$m$^2$.  Our best estimates for $I_{\mathrm{pk}}^{\mathrm{Im}}$ for $\Delta t=30$ fs are $(2.2 \pm 0.4,\ 3.9 \pm 0.6\ \mathrm{and}\  6.0 \pm 1.0) \times 10^{18}$ W/cm$^2$ from lowest to highest $U$.

\section{Discussion}

We compare $I_{\mathrm{pk}}^{\mathrm{RTS}}$ and $I_{\mathrm{pk}}^{\mathrm{Im}}$ graphically in Fig. \ref{fig:sidespectrumset-and-intensityvsenergy} (b), from which we can draw several conclusions.  First, given the uncertainties in $I_{\mathrm{pk}}^{\mathrm{Im}}$ and $I_{\mathrm{pk}}^{\mathrm{RTS}}$, the two are in general agreement.  Second, the uncertainty in $I_{\mathrm{pk}}^{\mathrm{Im}}$ is quite large ($\sim16.1\%$), with the largest contribution coming from the calibration of the camera ($\sim$15.7\%).  Consequently, a higher precision comparison is possible if an improved pixel area calibration is made.  It is interesting to note that the largest and smallest $I_{\mathrm{pk}}^{\mathrm{RTS}}$ datum points are in good statistical agreement with $I_{\mathrm{pk}}^{\mathrm{Im}}$($30$~fs), even ignoring the $A_{\mathrm{pix}}$ uncertainty; $I_{\mathrm{pk}}^{\mathrm{RTS}}$($U=2.67$ J$) > I_{\mathrm{pk}}^{\mathrm{Im}}$(30 fs) while $I_{\mathrm{pk}}^{\mathrm{RTS}}$($U=0.99$ J$) < I_{\mathrm{pk}}^{\mathrm{Im}}$($30$ fs).  At the same time, $I_{\mathrm{pk}}^{\mathrm{RTS}}$ ($U=1.73$ J) is lower than $I_{\mathrm{pk}}^{\mathrm{Im}}$($30$ fs) by more than $3\sigma$, and consistent with $I_{\mathrm{pk}}^{\mathrm{Im}}$($60$~fs).  There are three contributions to this observation.  First, as mentioned earlier, the spectra are contaminated with plasma recombination lines, making the determination of $\lambda^{(2)}_\mathrm{r}$ difficult; the observed onset wavelengths might be shifted to lower values.  Second, the $I_{\mathrm{pk}}^{\mathrm{RTS}}$ measurements were taken over several hours in the following order: $U=2.67$, $0.99$, and $1.73$ J.  The pulse width of this laser, which will affect the peak intensity, is known to drift with time; we had no capacity to monitor the pulse width in the target area in real time so the laser was not re-optimized after the initial optimization at the beginning of the day.  Third, changing the pulse energy typically requires re-optimization for peak performance.  Consequently, it is not surprising that the intensity was lower later in the day and after adjusting the energy.  To check for possible issues of this kind, we took a final $I_{\mathrm{pk}}^{\mathrm{RTS}}$ measurement at $U = 2.67$ J after the $U = 1.73$ J measurement, the green triangle in Fig. \ref{fig:sidespectrumset-and-intensityvsenergy} (b).  The fact that the green $U =2.67$ J point is lower than the black point is supporting evidence that the lower intensities after energy adjustment is likely caused by the laser not being optimized.  Evidently $I_{\mathrm{pk}}^{\mathrm{RTS}}$ is sensitive to changes in the beam properties leading to changes in the intensity, whereas $I_{\mathrm{pk}}^{\mathrm{Im}}$ is not.  While our results suggest the indirect image and RTS methods are in reasonable agreement, they also point out the need for real-time monitoring of the laser characteristics if the former is employed for determining the peak intensity.

Our study clearly shows that the RTS approach has promise as a diagnostic tool.  Nevertheless, there are issues that demand further investigation.  The most critical of these is a lack of knowledge of the conditions necessary to ensure that the $\lambda^{(n)}_{\mathrm{r}}$ observed is associated with Thomson scattering of electrons experiencing the peak intensity of the pulse.  Electrons born on axis near the peak of the pulse would seem to have the best chance of contributing to the RTS signal associated with the peak intensity. An electron born off axis with near zero velocity will not experience the peak intensity because (1) the ponderomotive energy ($U_p$) it picks up from the field will at best propel it no closer to the axis than its birth position and (2) the ponderomotive force tends to push it further and further away from the axis with each oscillation \cite{Gao:2006}.  Many, electrons fall into this category.  It is unclear if electrons born on the axis, but early in time before the pulse reaches its peak intensity, will remain close to the axis long enough to experience the peak intensity or be scattered away from the axis by the increasing ponderomotive force.   More subtle issues associated with the carrier phase and the time of birth of the electron also may play a role.  For example, electrons born either stationary at a peak of an optical cycle or with an initial velocity transverse to the laser axis at the \textit{right time} during a cycle so that it is turned around by the field, both have a chance to remain near the laser axis as they undergo Thomson scattering.  (Only the initial transverse velocity of the electron is important because the initial longitudinal velocity does not affect whether the electron is scattered away from the axis.) The dependence on phase has been discussed theoretically; see, for example, \cite{Gao:2004} and references therein.  To summarize, many scattered electrons either (1) leave the beam axis before the peak of the pulse envelope, never experiencing the peak intensity or leave the axis before the pulse is complete.  Either way many electrons may contribute Doppler-shifted components to the RTS spectra that are smaller than expected from a plane-wave analysis.  Consequently, $I_{\mathrm{pk}}^{\mathrm{RTS}}$ might only be a lower limit to the true intensity.  In order to understand these issues more fully, and to determine which RTS electrons properly reflect the peak intensity, both a numerical model and more experimental investigations are required.

\section{Conclusions}

The measurements reported in this manuscript were performed with N$_2$.  In addition to N$_2$ yielding ten electrons/molecule, the electrons are liberated at several different intensities, a few near the expected peak intensity.  This gives us confidence that at least a few electrons are present when the intensity is at its peak.  As argued above, electrons liberated near the peak intensity may have a better chance of experiencing the peak intensity.  Measuring the RTS signal for different gases with different ionization potentials and number of electrons liberated (e.g., He, Ar, N$_2$, etc.) could provide a better understanding of the role the ponderomotive force plays in the RTS, particularly for steep field gradients. It might also be possible to perform a pump/probe investigation to determine the time history of the electrons.  

Capturing the RTS signal at $(\theta, \phi) = (\pi/2,0)$ is most useful between $10^{18}$ and  $10^{19}$ W/cm$^2$.  At some point, which we estimate to be $\sim10^{19}$ W/cm$^2$, the harmonics will start to overlap, making it difficult to extract $\lambda^{(n)}_\mathrm{r}$ as we have done.  It is clear from inverting Eq.\ \ref{eq:intensity_ds} that the overlap between harmonics can be slowed by observing the radiation at smaller $\theta$.  While additional information about the RTS is contained in the angular distribution of the radiation, as suggested by Harvey \cite{Harvey:2018}, this approach will run into issues when $\theta$ gets too small, making it difficult to prevent the laser light from being captured with the RTS radiation.  We estimate that this will occur before $10^{21}$ W/cm$^2$.  A second possibility would be to pick off a calibrated portion of the energy and always measure the intensity between $10^{18}$ and $10^{19}$ W/cm$^2$.  A combination of picking off $\sim 1$ \% and capturing the radiation at smaller angles should allow intensities to be measured up to about 10$^{23}$ W/cm$^2$.  At this stage, quantum effects and radiation reaction will need to be considered.  A third approach is to monitor the infrared ($n = 1$) directly.  As the electrons become more relativistic, $\theta$ again becomes smaller.  Between 10$^{23}$ and 10$^{25}$ W/cm$^2$, it is not clear how to implement the red shifts with initially cold electrons to monitor the intensities.  The more complicated blue-shifts from higher-order Thomson scattering may need to be employed \cite{Yan:2017}.  As the intensity approaches $10^{25}$ W/cm$^2$, a new paradigm for characterizing the intensity based on proton RTS may open.  When an effective normalized vector potential for protons approaches unity (i.e., $a_p = e E \lambda / 2 \pi m_p c^2 \sim1$), the motion of free protons will become relativistic in the laser field and could emit radiation similar to that emitted by the electron, but for intensities about $(m_p/m_e)^2$ times larger than the electron case.  To our knowledge, this has never been discussed in the literature.

\section*{Appendix}
\def\theequation{A\arabic{equation}}
\setcounter{equation}{0}

In this appendix, we derive of Eq. 2.  We start by defining the intensity profile, which is a function of time and space in the focal plane,

\begin{equation}
	I(t,x,y) = I_0g(t)f(x,y),
\end{equation}

\noindent where $I_0$ is the peak intensity ($I_0 \equiv I(t_0,x_0,y_0)$, with $t_0$, $x_0$, and $y_0$ being the time and position of peak intensity respectively); $g(t)$ and $f(x,y)$ are respectively the temporal and spatial shape functions with $g(t_0) = f(x_0,y_0)=1$.  The total pulse energy on target is

\begin{equation}
	U\equiv \iiint I(t,x,y) dtdxdy = I_0 \int g(t) dt \iint f(x,y)dxdy.
\end{equation}

\noindent We define the time integral of $g(t)$ as an effective pulse width,

\begin{equation}
	T_{\mathrm{eff}} = \int g(t)dt,
\end{equation}

\noindent where $T_{\mathrm{eff}} \propto \Delta t\ (\mathrm{FWHM})$.  For a Gaussian shape ($g(t) = e^{-4\ln2(t/\Delta t)^2}$), for example, $T_{\mathrm{eff}} = \sqrt{\pi/\ln2} \Delta t/2 \approx 1.064 \Delta t$.   We next convert the spatial integral over $f(x,y)$ to a discrete sum,

\begin{equation}
	\iint f(x,y)dxdy \approx \sum_{i,j} f(x_i,y_j) \Delta x \Delta y,
\end{equation}

\noindent where $\Delta x \Delta y$ corresponds to $A_{\mathrm{pix}}$, the area in the focal plane captured by a pixel.  We identify $f(x_i,y_j)$ as the count at pixel $(x_i,y_j)$, divided by the peak count $C_{\mathrm{pk}}$.  We determined $C_{\mathrm{pk}}$ from the focal spot image by taking the set of nine pixels ($3\times3$ region) near the vicinity of the peak with the largest total count, and letting $C_{\mathrm{pk}}$ be equal to the average of the nine pixels.  Summing over all pixel comprising the pulse, we therefore obtain 

\begin{equation}
	\sum_{i,j}f(x_i,y_i) = \frac{C_{\mathrm{sum}}}{C_{\mathrm{pk}}},
\end{equation}

\noindent where $C_{\mathrm{sum}}$ is the sum of the counts of the image pixels comprising the pulse.  Thus, we have

\begin{equation}
U=I_0 T_{\mathrm{eff}} \frac{C_{\mathrm{sum}}}{C_{\mathrm{pk}}}A_{\mathrm{pix}}.
\end{equation}
Solving for $I_0=I_{\mathrm{pk}}^{\mathrm{Im}}$, we arrive at Eq. \ref{eq:image-intensity}.

\section*{Funding}
National Science Foundation (PHY1806584); Natural Sciences and Engineering Research Council of Canada (NSERC) (RGPIN-2014-05736); Spanish Ministerio de Ciencia, Innovacion y Universidades (PALMA Grant FIS2016-81056-R); LaserLab Europe IV (654148); Junta de Castilla y Le\'on Consolidated Research, Junta de Castilla y Leon (167).

\section*{Acknowledgements}
	We thank Enam Chowdhury for helpful discussion regarding proton RTS; Mauricio Rico, Diego de Luis, Juan Hernandez, Diego Arana and Jon Imanol Api\~naniz Aginako for technical help before and during the experiment; Cruz Mendez and the VEGA laser team for running the laser during the experiment; Emma McMillian, Max Szostak and Jacob Smith for help with the data analysis.



\bibliography{RTS_Intensity_He_Final}






\end{document}